# A Three-Dimensional Model of Residential Energy Consumer Archetypes for Local Energy Policy Design in the UK

Tao Zhang*, Peer-Olaf Siebers, Uwe Aickelin

Intelligent Modelling & Analysis Group, School of Computer Science

University of Nottingham

*Corresponding Author's Email: Tao.Zhang@nottingham.ac.uk

*Corresponding Author's Tel: +44 (0)115 8232517

**Abstract**

This paper reviews major studies in three traditional lines of research in residential energy consumption in the UK, i.e. economic/infrastructure, behaviour, and load profiling. Based on the review the paper proposes a three-dimensional model for archetyping residential energy consumers in the UK by considering property energy efficiency levels, the greenness of household behaviour of using energy, and the duration of property daytime occupancy. With the proposed model, eight archetypes of residential energy consumers in the UK have been identified. They are: pioneer greens, follower greens, concerned greens, home stayers, unconscientious wasters, regular wasters, daytime wasters, and disengaged wasters. Using a case study, these archetypes of residential energy consumers demonstrate the robustness of the 3-D model in aiding local energy policy/intervention design in the UK.

## 1. Introduction

In the UK, there are approximately 26 million houses. Totally, the residential sector accounted for 29% of the overall energy consumption in 2009 (ONS, 2009). The energy was consumed in various types of households for the purposes of space heating (58%), hot water (25%), cooking (3%), lighting and appliances (14%) (DECC, 2009). The energy consumption in the residential sector accounts for 26% of overall UK $CO_2$ emission (Swan et al, 2010). As the UK is a heavily urbanised industrial country, most of the residential houses are intensively distributed around municipalities/cities. Thus cities have a huge impact, for good or ill, on UK energy sustainability. Currently attention on energy policy making has been primarily paid at international/national level. For example, the UK government set a 2020 national target of cutting $CO_2$ emission by 34% from 1990 levels. Designing energy policy at local (e.g. city) level has always been neglected. The vast majority of UK cities have traditionally regarded energy as somebody else's problem. Hence cities not only have a lot of catching up to do but are also lacking in the knowledge, experience and tools needed to design effective local energy policies. Some cities (e.g. Leeds and London) now are aware of their important roles in the UK energy sustainability and actively seek knowledge, experience and tools for local energy design (Keirstead and Schulz, 2010). Residential sector is one of the key areas they would like to look at, as this sector traditionally in the UK can be directly influenced by local city councils via various local energy policies/interventions.

Looking at existing literature there are three lines of research focusing on residential energy consumption in the UK. A first line of research using statistical techniques to understand a key question: what are the factors influencing residential energy consumption (e.g. Shorrock, 2003; Baker and Rylatt, 2008; Kelly, 2010; Summerfield et al, 2010); a second line of research focuses on understanding residential energy consumers' pro-environment behaviour and more importantly, how changes in behaviour can lead to reduced household energy consumption (e.g. Mansouri et al., 1996; Wood and Newborough, 2003; Defra, 2007; Defra, 2008); and a third line of research looks at the load profiles of residential buildings and analysing the factors causing these different types of residential energy load curves (e.g. Newborough and Augood, 1999; Yao and Steemers, 2005). [A comprehensive review of techniques and models in studying domestic energy consumption can be found in Swan and Ugursal (2009)].

Residential energy consumption is a complex issue highly related to the physical attributes of the homes in which people live, the energy systems (e.g. electrical appliances) within these homes, and the occupying people's behaviour of using energy (Yao and Steemers, 2005; Swan et al, 2010). Sometimes economists also consider macro economic factors such fuel prices and inflation (Summerfield et al., 2010). These factors inter-related, and effective local energy policies for managing residential energy consumption rely on a comprehensive understanding of them. Whilst each line of previous studies contributes to a part of our understanding of residential energy consumption, currently there is a lack of a model which is inclusive of all these factors for local energy policy makers.

The objective of the paper is twofold. Firstly, motivated by a desire to comprehensively understand the complex issue of residential energy consumption, the paper reviews major previous studies on residential energy consumption in the UK, and proposes a conceptual model which integrates the factors extensively studied previously to archetype residential energy consumers in the UK. Secondly, based on the residential energy consumer archetypes developed by using the proposal conceptual model, the paper draws energy policy implications at local level in the UK.

The paper is organised as follows. The second section reviews and discusses the three lines of research in residential energy consumption in the UK, and proposes a conceptual model for archetyping residential energy consumers. The third section, based on the archetypes of residential energy consumers developed in Section2, draws local energy policy implications for local authorities in the UK. The fourth section discusses the proposed conceptual archetype model and local energy policy implications drawn from the model. The fifth section concludes the paper and refers to our future research.

## 2. Three-Dimensional Archetypes of UK Residential Energy Consumers in the UK

### 2.1 Brief Review of Residential Energy Research in the UK

Energy consumption in residential sector is caused by households. Traditionally, as described by Swan and Ugursal (2009), studies in household energy consumption usually adopt "top-down" and "bottom-up" approaches. Based on macroeconomic theories and the interactions between the energy sector and the whole economy, the top-down approach uses aggregated economic data such as empirically observed historic trends to predict future changes in energy consumption and $CO_2$ emissions. Top-down methods usually use econometrics and multiple linear regression models to explain the variance between dependent and independent variables (Swan and Ugursal, 2009; Kelly, 2011). Based on a disaggregated view, bottom-up methods estimate energy and emissions by using high resolution data combining factors such as physical, social, behavioural and demographic properties for a household, i.e. they see individual households as the basic units of domestic energy consumers and try to understand the patterns of energy consumption at that level (Swan and Ugursal, 2009; Kelly, 2011). Through a review of the literature in residential energy consumption in the UK, we have clearly identified three lines of research.

*Research Line 1: Understanding the factors influencing residential energy consumption in the UK*

In this line of research, apart from some economics literature giving descriptive statistics about residential energy consumption (e.g. ONS, 2009; DECC, 2009; Swan et al, 2010), studies try to find answers to this key question by using different statistical methods through both "top-down" and "bottom-up" approaches. Clearly, there are many factors that can influence households' energy consumption in the UK. These factors potentially include economic factors, seasonal factors, weather conditions, physical attributes of the properties, the energy systems within the properties, and the behaviour of the people living in the properties. There are some models using top-down regression methods to predict aggregate residential energy consumption in the UK (e.g. Utley and Shorrock, 2008; Summerfield et al, 2010). A widely adopted one was developed by Utley and Shorrock (2008) in the Building Research Establishment (BRE). This model uses domestic energy fact file (DEFF) to predict aggregate housing stock energy consumption. The equation of the model is:

$$Q = N*[97.84 + 2.18*(year - 1970) - 3.28 * T_e - 0.28 * \Delta H - 1.56*\Delta E\%]$$

Where:

- *Q* is the housing stock energy consumption
- *N* is the number of households (millions)
- *$T_e$* is the winter external temperature ($^0C$)

- *ΔH* is the improvement in the average dwelling heat loss relative to 1970 (W/$^0$C)

- *ΔE%* is the improvement in the average heating efficiency relative to 1970 (%)

In other words, the energy consumption of housing stocks is determined by *N, Te*, *ΔH* and *ΔE%.*

A recent improvement to the model is Summerfield et al (2010), which adds inflation adjusted energy price to the equation thus creating the annual delivered energy and price model (ADEPT). Although these top-down models appear to be robust in terms of predicting aggregate housing stock energy consumption in the UK residential sector, there are some obvious limitations with these models. For example, these models treat the UK housing stock homogenously with a very limited number of independent variables averaged over the UK on an annual basis. They fail to catch regional, local or individual household effects which may contain very important explanatory variables (e.g. fuel type, social-demographic and physical attributes of housing stocks in different areas, energy efficient technologies, and occupants' behaviour). Thus these models do not explain residential energy consumption in sufficient detail for policy making (Kelly, 2011; Kavgic et al, 2010).

More in line with the research presented in this paper are bottom-up methods, which look at the individual household level effects on household energy consumption by using high resolution data containing physical, social, demographic or sometimes even behavioural data of a household. The data requirement for bottom-up methods is significantly more demanding. Usually bottom-up methods require large quantitative datasets containing comprehensive specific attributes of households. Due to the demanding data requirements, there are a limited number of studies targeting the energy consumption in the UK residential sector. An early bottom-up study by Baker et al (1989) looks at the effects of household social-economic attributes (e.g. income, house size) on individual household demand for gas and electricity by using data over 50,000 households, pooled from 12 consecutive years of Family Expenditure Survey (1972-1983). A second bottom-up study by Baker and Rylatt (2008) develops an approach for improving the prediction of UK residential energy-demand by using data gathered from individual household questionnaire survey, supported by annual gas and electricity meter data and floor-area estimates derived from a geographic information system (GIS). The authors find that two variables, i.e. the number of bedrooms and regular home-working, have significant influence on the household energy consumption in the UK. A third bottom-up study by Kelly (2011) examines the causal relationships between explanatory factors and residential energy consumption by using structural equation modelling. Using the 1996 English House Condition Survey consisting of 2531 unique cases, the author finds that main drivers of residential energy consumption are the number of occupants living at home, household income, household heating patterns, living room temperature, floor area, and dwelling energy efficiency (i.e. the Standard Assessment Procedure (SAP) rate).

One notable point in this line of research is that SAP rate is found to have significant influence on household energy consumption (Swan et al, 2010; Kelly 2011). The Standard Assessment Procedure (SAP) is adopted by Government as the UK methodology for calculating the energy performance of dwellings. The calculation is based on the energy balance taking into account a range of physical factors that contribute to energy efficiency (BRE, 2009):

- materials used for construction of the dwelling
- thermal insulation of the building fabric

- ventilation characteristics of the dwelling and ventilation equipment
- efficiency and control of the heating system(s)
- solar gains through openings of the dwelling
- the fuel used to provide space and water heating, ventilation and lighting
- renewable energy technologies

The calculation is independent of factors related to the individual characteristics of the household occupying the dwelling when the rating is calculated, for example:

- household size and composition;
- ownership and efficiency of particular domestic electrical appliances;
- individual heating patterns and temperatures.

Ratings are not affected by the geographical location, so that a given dwelling has the same rating in all parts of the UK. The lasted version of SAP 2005, with scale ranging from 1 to 100, where 100 represents zero energy cost. It can be above 100 for dwellings that are net exporters. Based on SAP rates, each residential building can have an Energy Performance Certificate (EPC). Since October 2008, EPCs have been required every time a residential building is bought, sold or rented. EPCs are issued by accredited energy assessors alongside a supplementary report containing recommendations for property energy efficiency improvement. Energy assessors rate the energy performance of a property from "A" to "G", with "A" representing most energy efficient and "G" representing least energy efficient, as shown in Figure 1.

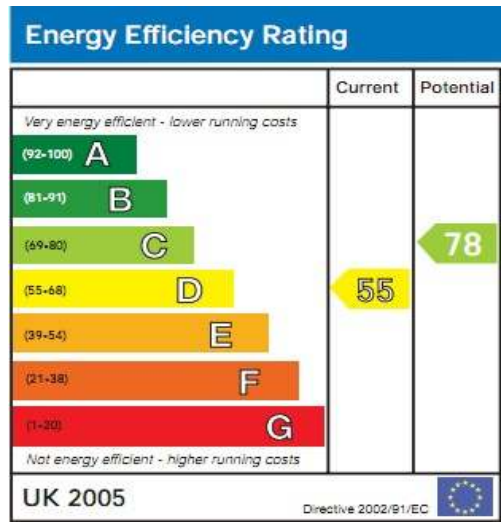

Figure 1: Energy Performance Certificate

In the first line of research in residential energy consumption, the "top-down" statistical models focus on understanding the causal relationship between aggregate variables/macro economic factors and the energy consumption in residential sector (e.g. Summerfield, 2010). The findings from these studies might be able to draw energy policy implications at macro level (e.g. national level). However, in terms of local energy policy design, they do not seem to be very useful. The "bottom-up" approach analyses the influence of micro level factors on household energy consumption. Many of these factors are the physical attributes of the properties (e.g. SAP model) and social-economic attributes of the households (e.g. family size and household income). Understanding how significantly these factors can influence household energy consumption might be useful for micro level (e.g. local level) energy policy design. However, a significant limitation of the approach is that it is not able to incorporate behavioural factors, which are known to be significant (Hitchcock, 1993) in the regression models. When considering designing local energy policies to influence household behaviour, these studies show little robustness.

*Research Line 2: Residential energy consumers' pro-environment behaviour*

Occupants' behaviour of using energy in a house can significantly influence the household's energy consumption. Recent studies in the UK have shown that energy use in households may vary by 2-3 times in properties with similar physical attributes (Summerfield et al, 2010b). Thus understanding residential energy consumers' behaviour of using energy is crucial in residential energy research. Occupants' behaviour of using energy is a complex issue and presents a big challenge for researchers. Looking at the literature, there is a line of research focusing on occupants' behaviour in the UK residential energy consumption studies. Mansouri et al. (1996) conduct a survey among householders resident in the south-east of England. The survey focused on identifying environmental attitudes and beliefs, energy-use behaviour, ownership levels for certain appliances and their utilisation patterns. Through the survey the authors find that members of the general public are (i) interested in receiving information concerning household energy use and the associated environmental impact, and (ii) willing to modify their behaviour in order to reduce household energy consumption and environmental damage. Therefore, the authors conclude that there is an urgent need to provide end-users with accurate energy-consumption and environmental-impact information, persuasively presented, to stimulate energy-rational and environmentally sustainable behaviour. Similar results were found in Brandon and Lewis (2002), in which the energy consumption of 120 households was monitored over a 9-month period. Participants in the study received feedback in various forms, i.e. consumption compared to previous consumption or to similar others; energy saving tips in leaflets or on a computer; or feedback relating to financial or environmental costs. From the study the authors find that participants with positive environmental attitudes, but who had not previously been engaged in many conservation actions, were more likely to change their consumption in response to the energy consumption information.

A more comprehensive study about residential energy consumers' behaviour is Defra (2007), which was a qualitative investigation into public understanding of sustainable energy consumption in the home. The investigation involved 12 focus groups of 8-10 people, energy audits and in-home advice (24 people), and depth interviews with audit participants (23 people). From the investigation Defra (2007) concludes a priority list of five possible goals: Priority 1 (joint): better energy management and usage in the home; Priority 1: (joint) installing insulation products; Priority 3: buying/installing energy efficient products/appliances; Priority 4: installing domestic micro-generation; Priority 5 (if at all): switch to a green energy tariff. This investigation later became part of Defra's broad study on people's pro-environmental behaviour (Defra, 2008). Defra (2008) covers people's broad pro-environmental behaviour, including behaviours in personal transport, home waste, home energy, home water, and eco-products. Based on people's willing to act and ability to act pro-environmentally, Defra (2008) classifies 7 segments of population (see Figure 2): positive greens, waste watchers, concerned consumers, sideline supporters, cautious participants, stalled starters, and honestly disengaged. Based on the segmentation, Defra (2008) draws some policy implication regarding encouraging people to act pro-environmentally.

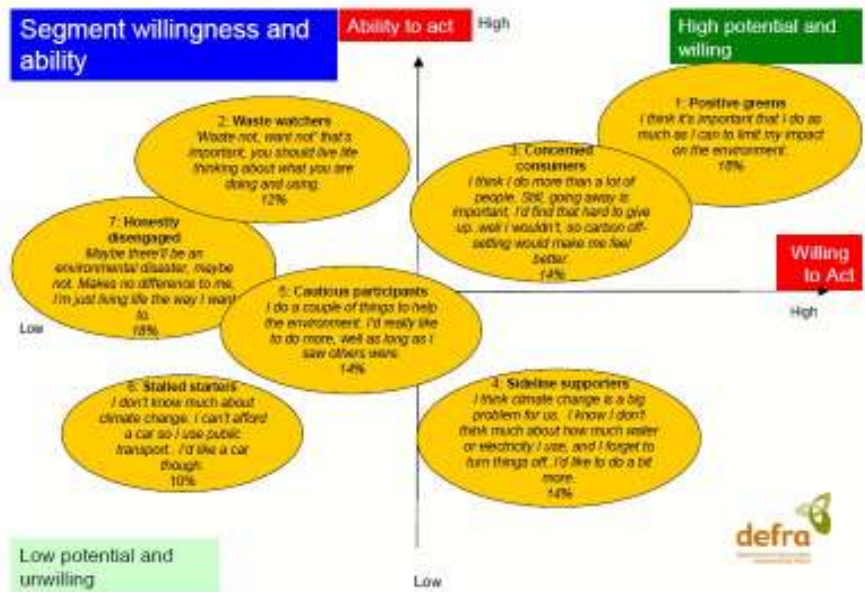

Figure 2: Seven Segments of Population Identified by Pro-environment Behaviour

(Defra, 2008)

In the second line of research in residential energy consumption, studies primarily focus on residential energy users' behaviour by using both quantitative and qualitative methods. Findings from these studies indicate that residential energy consumers are willing to change their behaviour in order to save money and become greener (Mansouri et al., 1996), and providing regular feedback of energy consumption at homes is an effective approach for changing residential energy consumers' behaviour (Hargreaves et al, 2010). These findings suggest that providing residential consumers with smart metering would be an effective energy policy. Defra (2008) even identifies seven population segments based on people's willingness and ability to act pro-environmentally. With each segment, it suggests different potentially effective intervention. The limitation of this line of research in residential energy research is that they ignore the effects of the physical attributes of the properties, (i.e. they assume that the physical attributes of the properties allow every household to take the energy interventions they suggested, which turns out to be unrealistic). Additionally, Defra (2008) is not specifically at household level. Instead it looks at the overall population. Although it can somewhat aid macro residential energy policy design, it does not give sufficient information for energy policy design at local level.

*Research Line 3: Load profiling*

Load profiling in energy research often appears in electrical engineering for electricity market management mechanism design (Stephenson and Paun, 2000), demand-side management (Newborough and Augood, 1999), and supporting renewable energy system deployment (Yao and Steemers, 2005). Many different methods, ranging from clustering methods to neural networks and data mining, have been adopted for energy load profiling (Chicco et al, 2003). In residential sector, detailed energy load profiles are "an important prerequisite for the accurate analysis of new low-carbon technologies and strategies, such as distributed generation and demand-side management" (Richardson et al., 2008, p.1560). The energy load of a household depends on various factors including physical attributes of the energy system (e.g. types, numbers and power of electrical appliances) and the occupancy pattern of the household (e.g. the number of occupants and whether they are at home and active) (Yao and Steemers, 2005; Richardson et al., 2008). Thus two central questions in load profiling for residential energy consumers are: (1) what do household energy load profiles look like? And (2) how do different factors cause these different types of residential energy load curves.

In the UK, a notable study about load profiling in the residential sector is Yao and Steemers (2005). In this study, the authors consider composition of household (i.e. size of households), occupancy patterns, energy-consumption of domestic appliances, and energy-consumption of domestic hot water and propose a simple method of formulating load profile (SMLP) for the UK residential buildings. Using this method, the authors formulate the energy load profile of a UK average household (i.e. in 2002, the average number of persons per household in the UK was 2.31) in five most common scenarios:

- Scenario 1: Unoccupied period is from 09.00 to 13.00. One of the occupants in this type of household may have a part-time job in the morning session.
- Scenario 2: Unoccupied period is from 09.00 to 18.00. The occupants in the house all have full-time job.
- Scenario 3: Unoccupied period is from 09.00 to 16.00. The family of this type of household may have a child to look after when school closed.
- Scenario 4: The house is occupied all the time. The family of this type of household may have minor child to look after or is of retired couples and single.
- Scenario 5: Unoccupied period is from 13.00 to 18.00. One of the occupants in this type of household may have a part time job in the afternoon session.

The load profiles in the five scenarios are shown in Figure 3.

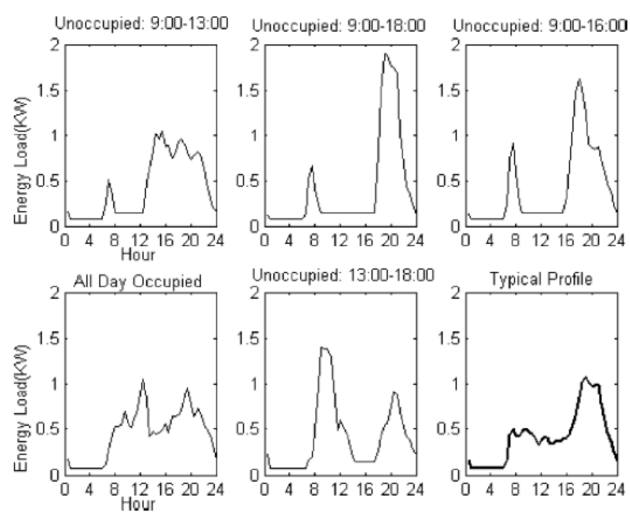

Figure 3: Energy Load Profiles of a UK Average Household (Yao and Steemers, 2005)

The method can be used to look at the energy load profiles at both micro level (e.g. an individual household) and macro level (e.g. a community). The authors also consider seasonal effects and topologies of properties in the UK (e.g. flat, semi-detached, detached, mid-terraced) when using the method for residential load profiling. One limitation of the method, as the authors pointed, is the lack of high resolution data about household occupancy pattern, which later was targeted in a study by Richardson et al. (2008). In that study, Richardson et al. develop a thorough and detailed method for generating realistic occupancy data for UK households, based upon surveyed time-use data describing what people do and when at homes. The model generates statistical occupancy time-series data at a ten-minute resolution and takes account of differences between weekdays and weekends. The model also indicates the number of occupants that are active within a house at a given time, which is important for example in order to model the sharing of energy use (shared use of appliances, etc.).

In the third line of research in residential energy consumption, studies look at the patterns of residential household energy load profiles, and how occupancy patterns, home energy systems, and property typologies cause different patterns of home energy load profiles. Studies in this line of research specifically focus on individual households. Findings from this line of research can provide useful information for energy supply tariff design and demand-side management. The limitation of the studies in this line of research is that they fail to consider behaviour change, i.e. they assume that when a house is occupied, the occupants are actively engaged in energy consumption. One reality however is that when people are at home, they can also change their behaviour of using energy from very active to less active or even inactive if a particular energy intervention (e.g. smart metering) works. Load profiling in residential energy research can aid demand-side energy interventions design at local level. However, if we were to develop more effective local energy policies/interventions by using demand-side techniques in conjunction with other techniques, we would need more information.

**2.2 The Proposed Three-Dimensional Archetype Model of Residential Energy Consumers in the UK**

The limitations of the traditional three lines of research in residential energy consumption in the UK bring us a desire to develop a conceptual model which integrates them for archetyping residential energy consumers in the UK. Since currently local authorities lack tools and knowledge for local energy policy/intervention design, this archetype model might help them with this. In the first line of research in residential energy consumption in the UK, many physical attributes of a property were examined to have big influence on the household energy consumption. Thus we consider a first dimension as energy efficiency level of the property. The energy efficiency level of a property is related to the physical attributes of the property, and independent of the number of occupants in the property and their behaviour. Drawing on the idea of the second line of research in residential energy consumption, we consider greenness of a household's behaviour of using energy as the second dimension. The third line of research in residential energy consumption in the UK shows that the energy load profile of a household is highly related to its length of daytime occupancy period. We thus consider the length of daytime occupancy period as the third dimension. If all each of the three dimensions has two measures, we can derive a three-dimensional model with eight archetypes of residential energy consumers, as shown in Figure 4. The eight archetypes of residential energy consumers are summarized in Table 1.

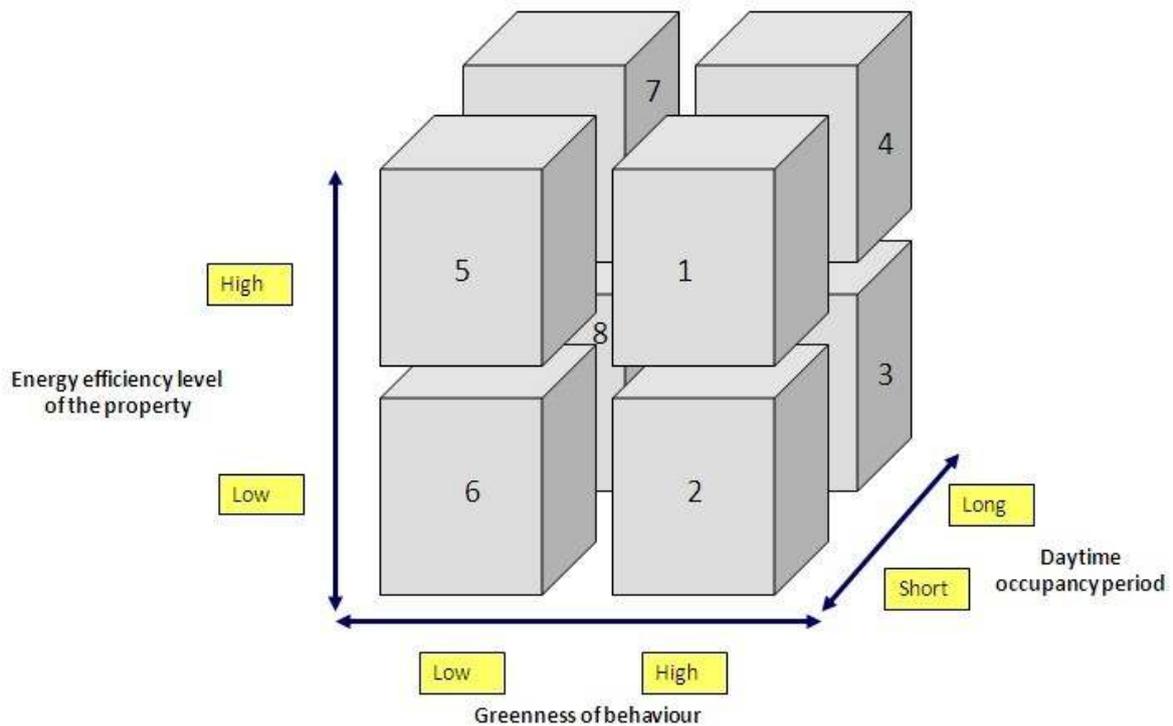

Figure 4: Three-Dimensional Archetype Model of Residential Energy Consumers in the UK

Table 1: Attributes of the UK Residential Energy Consumer Archetypes

| Archetype | Attributes | | |
|---|---|---|---|
| | **Property energy efficiency level** | **Greenness of behaviour** | **Duration of daytime occupancy** |
| 1: Pioneer Greens | High | High | Short |
| 2: Follower Greens | Low | High | Short |
| 3: Concerned Greens | Low | High | Long |
| 4: Home-Stayers | High | High | Long |
| 5: Unconscientious Wasters | High | Low | Short |
| 6: Regular Wasters | Low | Low | Short |
| 7: Daytime Wasters | High | Low | Long |
| 8: Disengaged Wasters | Low | Low | Long |

## 3. Implications for Local Energy Policy/Intervention Design: A Case Study

These archetypes can guide local energy policy/intervention design. In the UK, local energy policies/interventions are referred to as the policies/interventions that can be made directly by the local city councils to shape the energy systems at city level. Residential houses within a locality are directly influenced by the local city council. Traditionally, city councils in the UK are not actively engaged in energy policy/intervention making. However, in the recent climate of cutting emission and energy saving, many city councils are aware that they have important roles to play in that regard. Compared to national energy policies which are at strategic level, local energy policies/interventions are at operational level, as show in Figure 5. That means that the UK national strategic energy policies will devolve down to local level (e.g. cities) energy policies/interventions, and whether targets set in national energy policies can be achieved is to a large extent determined by the effectiveness of local energy policies/interventions.

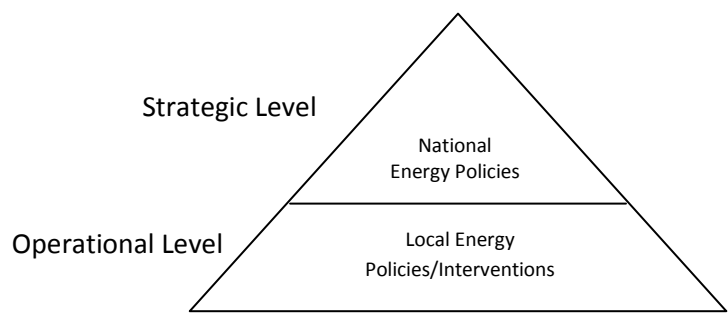

Figure 5: Levels of Energy Policies

In the residential sector, the effectiveness of a local energy policy/intervention is determined by whether the policy/intervention targets the right households. For example, installing smart meters in houses where the occupants do not make any sensible responses to the feedback of energy consumption will be an ineffective energy intervention. The 3-D archetype model of UK residential energy consumers can help us to design effective local energy policies/interventions.

Here in the paper we use a city as an example. The reason for using a city is that in the City Energy Future research project, we have investigated the possible local energy policies/interventions that city councils in the UK can take to manage the residential energy consumption, which are shown in Table 2.

Policies/interventions must target the right energy consumers in order to be effective. We can fit these energy policies/interventions into the eight archetypes developed by the 3-D model, and draw energy policy/intervention for city councils in the UK, as shown in Table 3.

Table 2: Potential City Level Energy Interventions

| Enhancement in end-use efficiency | Behavioural change by end-users |
|---|---|
| Technological:<br>• Free/means-tested/reduced cost measures for households<br>   o Insulation (loft, cavity, window etc.)<br>   o Smart meters<br>   o Boilers<br>   o Lighting<br>   o Domestic micro-generations<br>Social-Technical:<br>• Working with Housing Associations and other social landlords<br>• Variable council tax based on energy rating of property<br>• Education<br>   o Available energy tariffs<br>   o Behaviours<br>   o Energy awareness<br>   o Promoting use of external sources e.g. EST advice centre etc.<br>   o Loan energy readers | • Moral pressure<br>   o Green awards<br>   o Energy audits – quantitative league tables<br>   o Energy Bingo!<br>   o Use of social networking to highlight energy issues and opportunities<br>   o Campaigns and prioritisation thereof<br>• Accreditation scheme for private landlords (household and commercial properties)<br>• Working with estates agents to promote EPCs<br>• Facilitating access to existing support/funding/tech<br>   o Ease of access to info and applications for schemes – Strategic Energy Body<br>   o Ease of use of products of schemes<br>• Make energy efficiency easier than energy inefficiency<br>• Make it clear what the right thing is/impossible to do the wrong thing<br>   o Effective visual instruction<br>• Encourage community spirit<br>• Signing up to commitments/charters<br>• Small scale trials to wider roll-out |

Table 3: Energy Policy/Intervention Implications of the Eight Archetypes of Residential Energy Consumers in the UK

| Archetype | Enhancement in end-use efficiency | | Behavioural change by end-users |
|---|---|---|---|
| | Technological | Social-technical | |
| **1: Pioneer Greens** | Smart meters and domestic micro-generation | Set a lower council tax rate for the high energy efficiency of the properties | Give Green Awards for consuming less energy; Give more tips about how to saving energy at home; Encourage them to sign up variable energy tariffs; |
| | Households of this archetype consume least energy. Thus the energy intervention focus of households of Archetype 1 should be on maintaining their energy saving behaviour and exploring the possibilities of further energy saving. Households of Archetype 1 would be very interested in energy innovations such as smart metering and domestic micro-generations. Most of its energy consumption occurs in the evenings thus it would be useful to encourage the household to sign up variable energy supply tariffs. | | |
| **2: Follower Greens** | Insulations to improve property energy efficiency; Smart meters; | Set a higher council tax rate for the low energy efficiency of the property; Offer government sponsored property energy efficiency checks; | Work with estate agents or private landlords to promote EPCs; Encourage them to sign up variable energy tariffs; |
| | Due to the low energy efficiency of the property, households of this archetype consume more energy. Thus the energy intervention focus of households of Archetype 2 should be on improving their proper energy efficiency and maintaining their energy saving behaviour. It would be useful to help them check the energy efficiency level of the property, or have an EPCs issued by a professional assessor. If the local authority was to provide free home insulations or smart metering, they would be a nice target. Most of their energy consumption occurs in the evenings thus it would be useful to encourage the households to sign up variable energy supply tariffs. | | |
| **3: Concerned Greens** | Insulations to improve property energy efficiency; Smart meters; | Set a higher council tax rate for the low energy efficiency of the property; Offer government sponsored property energy efficiency checks. | Work with estate agents or private landlords to promote EPCs; Encourage them to sign up variable energy tariffs; |
| | Many occupants of households of this archetype are economically inactive and live in properties owned by the government or housing associations. Thus the council can provide government sponsored insulation services to improve the energy efficiency for these households. Also as the daytime occupancy of the properties is long, it is vital to provide regular feedback of energy consumption to the occupants through smart metering, and encourage them to sign up variable energy supply tariffs. | | |
| **4: Home-Stayers** | Smart meters and domestic micro-generation; | Set a lower council tax rate for the high energy efficiency of the properties | Give more tips about how to saving energy at home; Encourage them to sign up variable energy tariffs; |
| | Households of this archetype are similar to those of Archetype 1, but they have higher daytime energy consumption. Thus apart from the energy interventions applicable to Archetype 1, it is important to emphasize the effect of regular feedback of energy consumption through smart metering. This can help them dynamically manage their energy consumption. Households of this archetype can be the main targets of smart metering. | | |
| **5: Unconscientious Wasters** | | Enhance their energy awareness by providing educational programmes; Loan energy readers; | Energy audits—quantitative league tables; Use of social networking to highlight energy issues and opportunities; Encourage community spirit; Signing up to commitments/charters; |
| | Providing feedback of energy consumption to households of Archetype 5 would be less useful because of their low energy awareness. Thus the key intervention focus of Archetype 5 is on education and moral pressure. | | |
| **6: Regular Wasters** | Insulations to improve property energy efficiency; | Set a higher council tax rate for the low energy efficiency of the property; Offer government sponsored property energy efficiency checks; Enhance their energy awareness by providing educational programmes; Loan energy readers; | Energy audits—quantitative league tables; Use of social networking to highlight energy issues and opportunities; Encourage community spirit; Signing up to commitments/charters; |
| | Similar to those in Archetype 5, households in Archetype 6 is of low energy awareness thus providing feedback of energy consumption to them through smart metering in the initial stage is less useful. Thus the key energy intervention focus of Archetype 6 should be on improving property energy efficiency through insulations and enhancing their energy awareness through education and moral pressure. | | |
| **7: Daytime Wasters** | | Enhance their energy awareness by providing educational programmes; Loan energy readers; Work with property management companies to introduce incentives to encourage energy saving. | Energy audits—quantitative league tables; Use of social networking to highlight energy issues and opportunities; Encourage community spirit; Signing up to commitments/charters; |
| | Similar to the households in Archetype 5, the key intervention focus of Archetype 5 is on education and moral pressure. | | |
| **8: Disengaged Wasters** | Mandate Insulations to improve property energy efficiency | Set a higher council tax rate for the low energy efficiency of the property; Offer government sponsored property energy efficiency checks; Enhance their energy awareness by providing educational programmes; Mandate the requirement of EPCs | Energy audits—quantitative league tables; Use of social networking to highlight energy issues and opportunities; Encourage community spirit; Signing up to commitments/charters; |
| | Households of Archetype 8 consume the most energy. Thus the centre of energy intervention focus of Archetype 8 should be on mandating insulations to improve the energy efficiency of their homes, and enhancing their energy awareness through education and moral pressure. | | |

## 4. Discussion

The proposed conceptual model has several advantages. Firstly, it integrates the factors that have been extensively studied in previous research in residential energy consumption in the UK. This could help us to avoid drawing skewed energy policies/interventions from one single dimension. Secondly, it is specifically located at household level thus could be a powerful tool for local energy intervention design in residential sector. Thirdly, by applying more concrete measures/scales on each dimension based high resolution data about local households, city councils can derive more specific archetypes of local residential energy consumers, with each archetypes referring to more specific and effective energy interventions. Although this requires high resolution data about local households, local authorities can somehow manage to gather it—they are those who understand their local communities most.

As the model is a conceptual one for archetyping residential energy consumers in the UK, it is important to acknowledge its limitations. Firstly, this model does not consider home appliances, which are usually considered to be an important factor influencing household energy consumption (Yao and Steemers, 2005). Having what electrical appliances at home is the decision made by the household, and it is usually assumed that a household in the UK has a standard set of home electrical appliances (Yao and Steemers, 2005). Thus ignoring electrical appliances would not cause significant adverse effects on the local energy policy/intervention implications from the model. A second limitation is that, while property energy efficiency level is independent of occupants' behaviour and length of daytime occupancy, there might be correlations between the later two. However, in residential energy research these are no studies indicating this. A third potential limitation is that the applicability of the conceptual model is within the UK. This is because the bases for development the model—the three lines of research in residential energy consumption—are in the UK context. We do not presumptuously claim that the model is applicable in other areas, as different areas may have very different situations about energy consumption in the residential sector.

## 5. Conclusions and Future Research

The paper reviews three traditional lines of research in residential energy consumption in the UK, and based on the review it proposes a three-dimensional models for achetyping residential energy consumers in the UK. With the model, eight archetypes of residential energy consumers in the UK have been identified. They are: pioneers greens, follower greens, concerned greens, home stayers, unconscientious wasters, regular wasters, daytime wasters, and disengaged wasters. With these archetypes of residential energy consumers, the 3-D model demonstrates how it can aid local energy policy/intervention design in the UK cities. From the study in the paper we conclude that, in order to be effective, local energy policies/interventions rely on specific and concrete residential energy consumer archetypes, and using three dimensions, i.e. property energy efficiency level, greenness of behaviour, and length of daytime occupancy, to archetype residential energy consumers is an effective way for local energy policy/intervention design in the UK.

We will pursue future research with the model in two directions. The first direction, as mentioned before, is to incorporate more measures in each dimensions for more specific local energy/intervention design. We are now carrying out empirical surveys in a city in the UK to gather high resolution data about residential energy consumers. Hopefully the high resolution data will enable us to make more specific and effective local energy policy/interventions in the UK.

The second direction of future research is using the model to archetype residential energy consumers for energy research with computational simulation methods. In recent year, there has been a trend of using agent-based simulation in residential energy research area (e.g. Keirstead, 2006; Faber, et al, 2010; Zhang and Nuttall, 2011). The modelling rationale of agent-based simulation in residential energy research areas is that we model residential energy consumers as intelligent agents—residential energy consumer agents, which behave and interact in a virtual environment based on the various attributes of their real world counterparts. By applying different energy policies/interventions into the virtual environment, we can observe and analyse its system-level phenomena, from which we can draw policy implications.

Developing high fidelity residential energy consumer agents depends heavily on the archetypes of residential energy consumers. A generic residential energy consumer agent template requires a comprehensive understanding of the attributes of residential energy consumers from various dimensions. For the convenience of using agent-based simulation in broad energy research in the UK, we would like to extend the residential energy consumer archetypes based on the 3-D model to develop a high fidelity generic residential energy consumer agent template. More importantly, with the model we would like to investigate how we can enable a residential energy consumer transfer from a low energy-efficient archetype to a high energy-efficient archetype. This would be a very useful tool to aid local authorities in the UK with local energy policy/intervention design.

**Acknowledgements**

The work was financially sponsored by the UK Engineering and Physical Sciences Research Council (EPSRC), under the *Future Energy Decision Making for Cities - Can Complexity Science Rise to the Challenge?* project (EPSRC Grant References: EP/G05956X/1 and EP/G059780/1). The authors would like to thank their colleagues (particularly C. Bale, N. McClullen, W. Gale and T. Foxon) in the University of Leeds for their generous help with the research. The authors also acknowledge the help from the project partner Leeds City Council.